\documentclass[prb,groupedaddress,showpacs]{revtex4}
\usepackage{amssymb}

\usepackage{graphicx}
\usepackage{dcolumn}
\usepackage{bm}
\usepackage{amsmath}
\usepackage{color}
\newcommand{\beq}{\begin{equation}}
\newcommand{\eeq}{\end{equation}}

\begin{document}
\begin{sloppypar}
	
\end{sloppypar}
\title{Dielectric function and plasmons in graphene} 

\author{Antonio Hill}
\author{S. A. Mikhailov}
\author{Klaus Ziegler}
\affiliation{Institut f\"ur Physik, Universit\"at Augsburg, D-86135 Augsburg, Germany}

%\affiliation{Theoretical Physics II, Institute of Physics, University of Augsburg, D-86135 Augsburg,
%Germany}

%\affiliation{Theoretical Physics II, Institute of Physics, University of Augsburg, D-86135 Augsburg,
%Germany} 

\date{\today }

\begin{abstract}
The electromagnetic response of graphene, expressed by the dielectric function, 
and the spectrum of collective excitations are studied as a function of wave 
vector and frequency. Our calculation is based on the full band structure,
calculated within the tight-binding approximation. As a result, we find 
plasmons whose dispersion is similar to that obtained in the single-valley 
approximation by Dirac fermions. In contrast to the latter, however, we find 
a stronger damping of the plasmon modes due to inter-band absorption. Our 
calculation also reveals effects due to deviations from the linear Dirac 
spectrum as we increase the Fermi energy, indicating an anisotropic behavior 
with respect to the wave vector of the external electromagnetic field.
\end{abstract}

\maketitle

\section{Introduction}

Graphene, a single layer of carbon atoms arranged as a honeycomb lattice, is a semimetal
with remarkable transport properties \cite{geim1}. This is due to the band structure of the
material which consists of two bands which touch each other at two nodes. The electronic
spectrum around these two nodes is linear and can be approximated by Dirac cones \cite{geim1}. Many
unusal transport properties are controlled by the fact that the Fermi energy is at the nodes, 
where the density of states vanishes. However, graphene has been gated such that the Fermi 
energy can be freely tuned. This has opened a wide field for experiments \cite{geim2}.
In the following we will discuss collective excitation of the 2D electron gas in graphene,
caused by an external electromagnetic field. Although this is a standard problem in
semiconductor physics, it was studied in the case of graphene only in the Dirac approximation
around the nodes \cite{wunsch,sarma,macdonald}. 

% {\it dielectric function:}
We consider an electron gas which is subject to an external potential $V_i(\textbf{q},\omega)$. The response of 
the electron gas is a screening potential $V_s(\textbf{q},\omega)$ which is caused by the rearrangement of the 
electrons due to the external potential. Therefore, the total potential, acting on the electrons, is
\beq
V(\textbf{q},\omega)=V_i(\textbf{q},\omega)+V_s(\textbf{q},\omega) \ .
\eeq
$V_s$ can be evaluated self-consistently \cite{ehrenreich59} and is expressed via the
dielectric function $\epsilon(\textbf{q},\omega)$. Then the total potential reads \cite{mahan}
\beq
V(\textbf{q},\omega)=\frac{1}{\epsilon(\textbf{q},\omega)}V_i(\textbf{q},\omega) \ .
\eeq
The dielectric function is determined by the specific Hamiltonian of the electron gas. In the
present case this is the tight-binding Hamiltonian on a honeycomb lattice.

The aim of our study is to compare the dielectric function of the full honeycomb lattice with
previous studies of the single Dirac cone. In contrast to the latter, we can not rely on a
closed expression for the integrals but have to integrate over the Brillouin zone numerically.

\section{Model}

Using the tight-binding approximation, (quasi-) electrons in graphene are described by the
Hamiltonian \cite{ziegler06}
\beq
H=h_1\sigma_1+h_2\sigma_2
\eeq
with Fourier components
\beq
h_1=-t\sum_{j=1}^3\cos(\textbf{b}_j\cdot \textbf{k}), \ \ \ h_2=-t\sum_{j=1}^3\sin(\textbf{b}_j\cdot \textbf{k}) \ .
\eeq
$\sigma_{1,2}$ are the Pauli matrices and $\textbf{b}_{1,2,3}$ are the next neighbor vectors on the honeycomb lattice:
\beq
\textbf{b}_1=d(\sqrt{3}/6 , 1), \ \ \textbf{b}_2=d(\sqrt{3}/6 , -1), \ \ \textbf{b}_3=d(-1/\sqrt{3},0) \ ,
%b_1=d/2(1/\sqrt{3},1), \ \ b_2=d/2(1/\sqrt{3},-1), \ \ b_3=d(-1/\sqrt{3},0) \ .
\eeq
$t$ is the hopping parameter ($\approx 2.8eV$)
%The unit vectors are
%\[
% a_1 = d/2(\sqrt{3},-1), \ \ \ a_2=d(0,1), \ \ \ b=d/2(1/\sqrt{3},1),
%\]
and $d$ is the lattice constant ($\approx 1.42$ \AA).
The Hamiltonian $H$ satisfies the eigenvalue equation
$H|\textbf{k}l\rangle =E_{\textbf{k}l}|\textbf{k}l\rangle$ with eigenvalues $E_{\textbf{k}l}=(-1)^lE_\textbf{k}$ and the positive square root $E_\textbf{k}=\sqrt{h_1^2+h_2^2}$. Thus the index $l$ refers either to the conduction band ($l=2$) or 
to the valence band ($l=1$).

\section{Dielectric function}

The dielectric function can be calculated from the Lindhard formula. Assuming that the wave length of the electromagnetic wave is much larger than the lattice spacing, the longitudinal component reads \cite{ehrenreich59} 
\begin{equation}
\epsilon(\textbf{q},\omega)= 1-\frac{2 \pi e^2 }{q} \chi(\textbf{q}, \omega) \, ,
\label{dielectric}
\end{equation}
where
%\begin{align}
\begin{equation}
\chi(\textbf{q}, \omega) = \lim_{\delta \rightarrow 0} \sum_{\textbf{k}ll'} 
\frac{f(E_{\textbf{k},l})-f(E_{\textbf{k}'l'})}{E_{\textbf{k}l} - E_{\textbf{k}'l'}+\hbar \omega + i\hbar \delta}\\
%\times 
|\langle \textbf{k}'l'|e^{i \textbf{q} \cdot r} |\textbf{k}l\rangle |^2 \, 
\label{polariz}
\end{equation}
%\end{align}
with $\textbf{\textbf{k}}'=\textbf{\textbf{k}}+\textbf{q}$. $f(E)=1/(e^{\beta (E-\mu)}+1)$ is the Fermi-Dirac distribution function at inverse temperature $\beta=1/k_B T$
with chemical potential $\mu$.
In order to determine $\chi(\textbf{q}, \omega)$ we obtain, after some straightforward calculations for Eq. \eqref{polariz}, 
an expression for the polarizability ($\hbar=1$) as
%\begin{align}
%\chi(q, \omega)= \int_{B.Z.} \frac{d^2 k}{\Omega_{B.Z.}} \bigg{(}
%\nonumber f(E_k)\left\{ \frac{|\kappa_k^*\kappa_{k'} \pm 1|^2}{E_{k} \mp E_{k+q}+ \omega - i \delta}  \right\} \\
%\nonumber +f(-E_k)\left\{ \frac{|\kappa_k^*\kappa_{k'} \pm 1|^2}{-E_{k} \pm E_{k+q}+ \omega - i \delta} \right\} \\
%\nonumber +f(E_k)\left\{ \frac{|\kappa_k^*\kappa_{k'} \pm 1|^2}{E_{k} \mp E_{k+q}- \omega + i \delta} \right\} \\
%+f(-E_k)\left\{ \frac{|\kappa_k^*\kappa_{k'} \pm 1|^2}{-E_{k} \pm E_{k+q}- \omega + i \delta} \right\} \bigg{)} \, ,
%\label{integrand}
%\end{align}
%or
\beq
\chi(\textbf{q}, \omega)=\lim_{\delta \rightarrow 0} \sum_{s=\pm 1}\int_{BZ} |\kappa_\textbf{k}^*\kappa_{\textbf{k}+\textbf{q}} +s|^2
\bigg{(}\frac{f(E_\textbf{k})-f(-E_\textbf{k})}{E_{\textbf{k}} -s E_{\textbf{k}+\textbf{q}}+ \omega - i \delta}
+ \frac{f(E_\textbf{k})-f(-E_\textbf{k})}{E_{\textbf{k}} -s E_{\textbf{k}+\textbf{q}}- \omega + i \delta} 
\bigg{)}\frac{d^2 k}{\Omega_{BZ}} \ ,
\label{integrand}
\eeq
where $\kappa_\textbf{k} =(h_1-ih_2)/E_\textbf{k}$. $\Omega_{BZ}$ is the area of the two-dimensional Brillouin zone
$\Omega_{BZ}=\int_{BZ}{d^2 k}$. The index $s$ describes intraband (for $s=1$) and interband scattering (for $s=-1$).
The integral of Eq. %\eqref{dielectric} and 
\eqref{integrand} is evaluated numerically
for different values of the chemical potential $\mu$ and different directions of the wave vector $\textbf{q}$, using the limit
of zero temperature.
\begin{figure}
\begin{center}
	\includegraphics[scale=0.9]{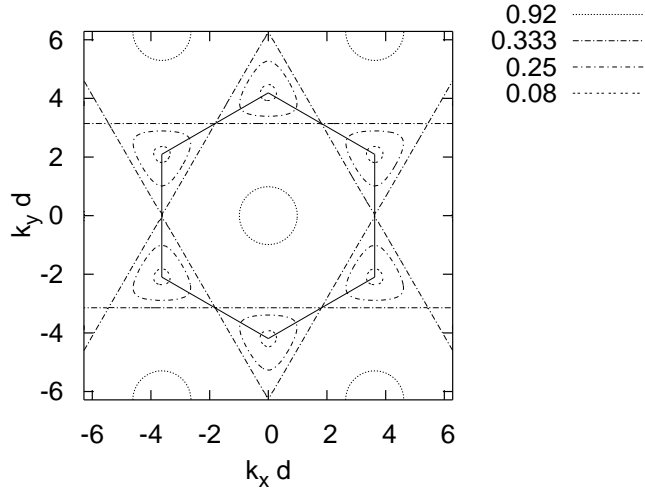} 
\hbox{ }\hbox{ }
\end{center}
\caption{Contourplot of the electronic dispersion for different energies. The Brillouin zone is indicated by the hexagon 
which connects the Dirac nodes. }
\label{bandstructure}
\end{figure}

\section{Plasmons}

Poles in $\omega$ of the inverse longitudinal dielectric function $1/\epsilon(\textbf{q},\omega)$ for a given wave vector $\textbf{q}$
correspond to collective excitations of electrons which are called plasmons.
These poles are located either on the real axis or in the complex plane away from the real axis. The latter can be considered as
damped plasmons, generated by scattering with individual electrons.
%There is a scattering rate $\eta$ which is related to the imaginary part of $\epsilon$. 
An imaginary term can appear in the
integral $\chi(\textbf{q}, \omega)$ of Eq. ({\ref{integrand}) if the denominator $E_{\textbf{k}} -s E_{\textbf{k}+\textbf{q}}+ \omega$ vanishes 
inside the Brillouin zone. In other words, if $(\textbf{q},\omega)$ is inside 
the band which is produced by the spectrum of the electrons, i.e. where an electronic wave vector $\textbf{k}$ exists that satisfies
\begin{equation}
E_{\textbf{k}+\textbf{q}}- sE_\textbf{k}=\omega\ \ \ (s=\pm1) \ ,
\label{qpspectrum}
\end{equation}
(cf. Fig. 1), scattering between plasmons and electrons is possible and will lead to damping of plasmons. 
On the other hand, outside the spectrum of electrons (i.e., when there is no electron wave vector $k$ which 
solves Eq. (\ref{qpspectrum})) we obtain undamped plasmons.

As an approximation we can expand the electronic dispersion around the nodes in the $E=0$ plane.
This gives two independent Dirac cones (valleys) with linear dispersion $E_k\sim \gamma k$ around each node. 
Then the poles of the inverse dielectric function can be evaluated exactly \cite{sarma,wunsch}. In this case the 
plasmon dispersion follows a square root: $\omega_P\sim q^{1/2}$. 

Collective excitations, on the other hand, depend on the spectral properties of the electrons. 
Therefore, deviations from Dirac cones can affect them.
We found that these deviations lead to a stronger damping of the electrons, since electronic excitations require
lower energies on the honeycomb lattice in comparison with the linearized (Dirac) spectrum. 
 
In the following we study the imaginary part of the inverse dielectric function\cite{macdonald}
\begin{equation}
 Im\left( \frac{1}{\epsilon(\textbf{q},\omega)} \right) = \frac{-\epsilon''}{\epsilon'^2 + \epsilon''^2} \,  ,
\end{equation}
where $\epsilon'$ ($\epsilon''$) is the real (imaginary) part of the dielectric function itself.
This quantity becomes a sharp Dirac delta function for a pole of $1/\epsilon(\textbf{q},\omega)$ 
on the real axis. If the pole is away from the real axis it becomes a Lorentzian which has 
a width $\epsilon''$, implying that the width of the Lorentzian is a measure for damping by 
electron scattering.

\section{Discussion}

The polarizability $\chi$ of Eq. (\ref{integrand}) is evaluated numerically and the corresponding
dielectric function is obtained from Eq. (\ref{dielectric}) for different values of the frequency $\omega$,
the wave vector $\textbf{q}$ and chemical potential (Fermi energy) $\mu$.
All energies are measured in units of the electronic bandwidth ($\Delta=3t$), and
wave vectors are measured in units of the inverse lattice constant.

The electronic dispersion is plotted for several Fermi energies in Fig. \ref{bandstructure}.
The contours are Fermi surfaces that indicate at which wave vectors electronic intraband scattering occurs for 
a given Fermi energy.
For $\mu=0.08$ we have intraband scattering only very close to the Dirac points and the dispersion is Dirac like.
As we increase $\mu$ we see the effect of ``warping'' and for $\mu=1/3$ intraband scattering between different
valleys is very likely because we have a connected Fermi surface. 
Finally, for $\mu=0.92$ we have only a circular Fermi surface around $k=0$ with parabolic
dispersion. This case corresponds with a conventional 2D electron gas.
% Apparently, we can tune our 2D system from a Dirac-like electron gas to a conventional parabolic electron gas by changing the Fermi energy. The latter can be in principle done for gated graphene \cite{geim1}.    

% {\bf List of plots with data and figure numbers in Table} \ref{table1}.

The plots in Fig. \ref{directions} show the anisotropy of the plasmon dispersion as a consequence of the
band structure of graphene. Remarkable is the strong deviation from the Dirac case for $q_x=q_y$ in
Fig. \ref{directions}c.
In Fig. \ref{energies} it can be seen how the plasmon dispersion changes with the Fermi energy $\mu$. The curvature
of $\omega_P(\textbf{q})$ is negative for low energies (a square-root behavior close to the Dirac points), 
it increases with $\mu$ and becomes positive for energies higher and larger momenta $q$. This behavior indicates 
a cross-over from the square-root law
of Dirac fermions to the behavior of the conventional 2D electron gas \cite{stern67}
\beq
 \omega_p(q)={\frac {\sqrt { \left( 4 a+v_F^2 q \right) q \left(q^4 v_F ^4 +4 
q^3 v_F^2 a + 16 k_F^2 a^2
 \right) } \left( v_F^ 2 q+2 a \right) }{4 \left( 4 a+ v_F^2 q \right) a k_F}}
\label{conveg}
\eeq
with $a=2ne^2/m$, the effective electron mass $m$, and the Fermi velocity $v_F$. 
Expansion for small $q$ gives
\beq
 \omega_p^2 \approx  a q + \frac{3}{4} v_F^2 q^2 \ ,
\eeq
which is compared with the findings for graphene at $\mu=0.91$ (cf. Fig. \ref{energies}c).
%\[
% \omega_p(q) 
%= \frac{\sqrt{a_0(2+qa_0)q(a_0^2 q^4 + 2a_0 q^3 + 4 k_F^2)}(v_F 
%a_0 q +v_F)}{2a_0k_F(2+qa_0)} \ ,
%\]
%where $a_0=4 \pi \varepsilon_0 \hbar^2/m_e e^2$ is the Bohr radius with the effective elctron mass $m_e$
%and $v_F$ the Fermi velocity.
Thus, electrons in graphene behave like in a conventional 2D electron gas.
This means that if it is possible to vary the Fermi energy between the linear and
the parabolic regime, the behavior of the fermions in graphene could be 'switched' from relativistic Dirac 
fermions to ordinary electrons.

In the theory presented above we have made two approximations. First, we have assumed that the wave functions $\phi_0({\bf r})$ of the electrons localized at each atom satisfy the condition $|\phi_0({\bf r})|^2\approx \delta({\bf r})$, i.e. we have ignored the spatial scale $a_0$ of $\phi_0({\bf r})$. Second, using the Lindhard formulas (\ref{dielectric})--(\ref{polariz}) we have ignored the local field effects \cite{adler62}. Both assumptions restrict the absolute value of the wave-vector $q$ by the conditions $qa_0\lesssim 1$ and $qd\ll 4\pi/3$, therefore the plasmons spectra are plotted in Figures \ref{directions} and \ref{energies} at $qd\le 1.5$, which is still within the applicability conditions of the theory. 

In Figure \ref{largeq}, however, we also present our test results for the plasmon spectrum at larger values of $q$, up to $qd\simeq 4$. In spite of at so large $q$ the theory is not quantitatively applicable, one can expect that it provides a reasonable qualitative description of the plasmon spectra in graphene. One sees that, while the low-frequency plasmon mode with the frequency $\omega_P\sim \sqrt{q}$ gets a large damping and disappears at $\omega/\Delta\simeq 0.5$,  at even higher energies $\omega/\Delta\gtrsim 0.8$ the plasmon mode reappears again. The both plasmon branches look like two parts of the same dispersion curve, ``teared'' to the low- and high-frequency pieces by the area of the substantial inter-band damping at $0.5\lesssim\omega/\Delta\lesssim 0.8$. This interesting result needs to be studied further within a more advanced theory which takes into account the local field effects \cite{adler62}.

%Moreover, we see a second resonance at higher energies and larger wave vectors in Fig. \ref{largeq}.

%At this point we have to emphasize the approximations we made in this model. As a consequence of the tight-binding approximation the wave functions are localized at each atom so they are $\delta$-function like. In a more realistic approach the wave functions should have a finite width with exponential decay. This results in a prefactor for the matrix elements which decreases with growing $\textbf{q}$, in our case this prefactor is $1$ for all values of $\textbf{q}$. Another fact is, that we did not include the periodicity of $\textbf{q}$ in our matrix elements. If we want to compare our results with real experiments both assumptions restrict $\textbf{q}$ to values much smaller than the length of the reciprocal lattice vector. The remaining question is how the second resonance will change (especially for $qd>2$) if local field corrections \cite{adler62} are included.

\begin{figure}[ht]
\begin{center}
	\includegraphics[scale=0.83]{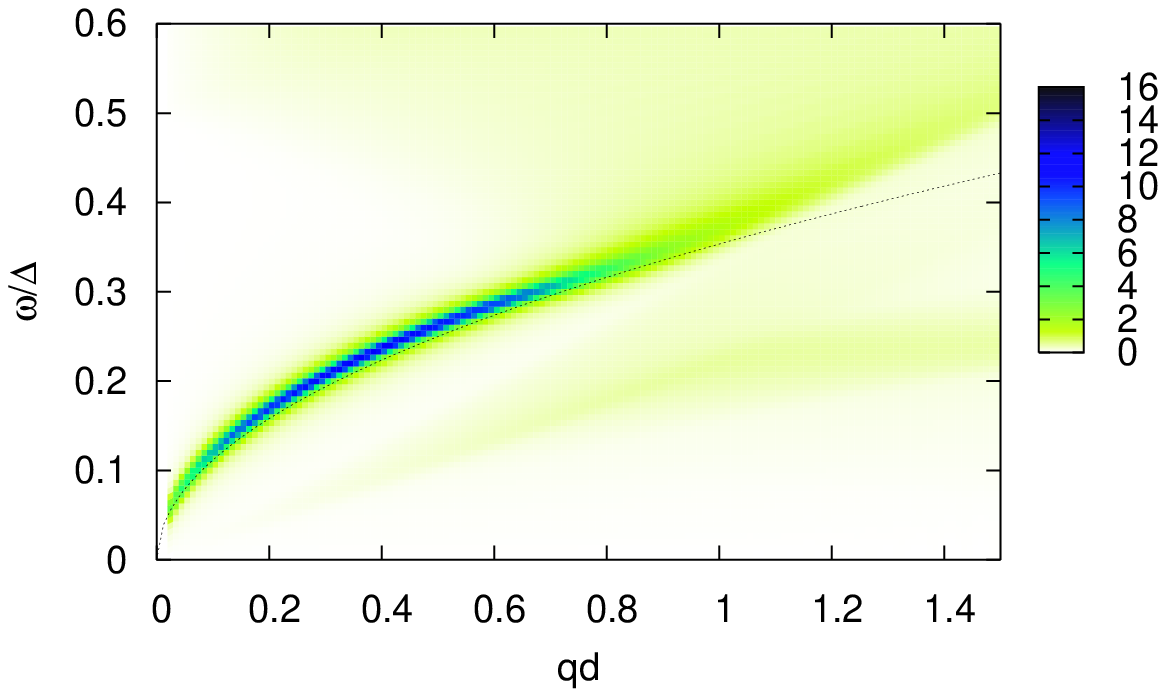}
	\includegraphics[scale=0.83]{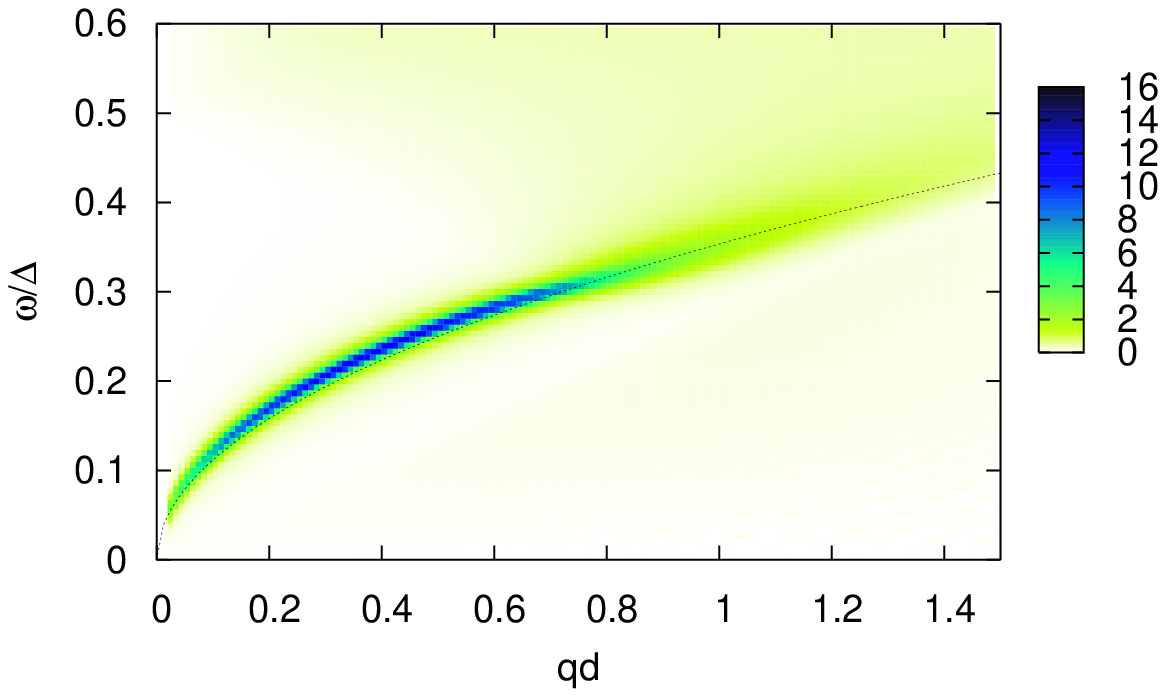}
	\includegraphics[scale=0.83]{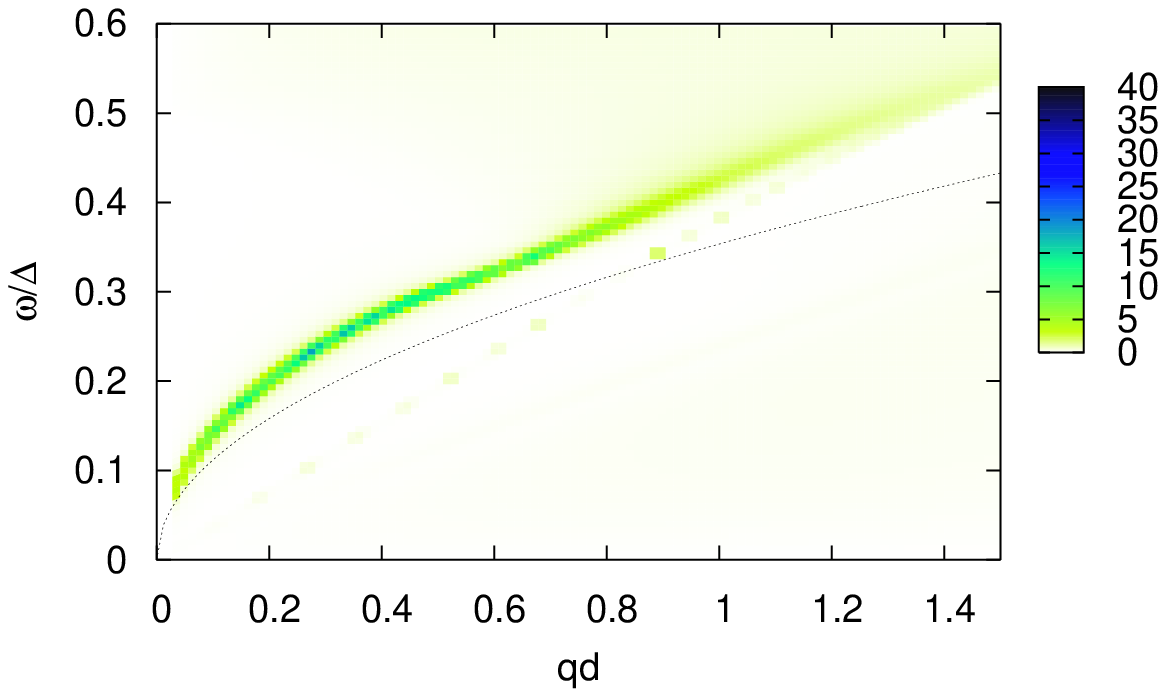}
\end{center}
\caption{(Color online) Plasmon dispersion for $\mu=0.25$ and different directions of the $q$ vector
a) $q_x=0$, b)$q_y=0$, c) $q_y=q_x$. The square-root behavior of the Dirac case
is also shown as a dashed curve.}
\label{directions}
\end{figure}

\begin{figure}[ht]
\begin{center}
	\includegraphics[scale=0.83]{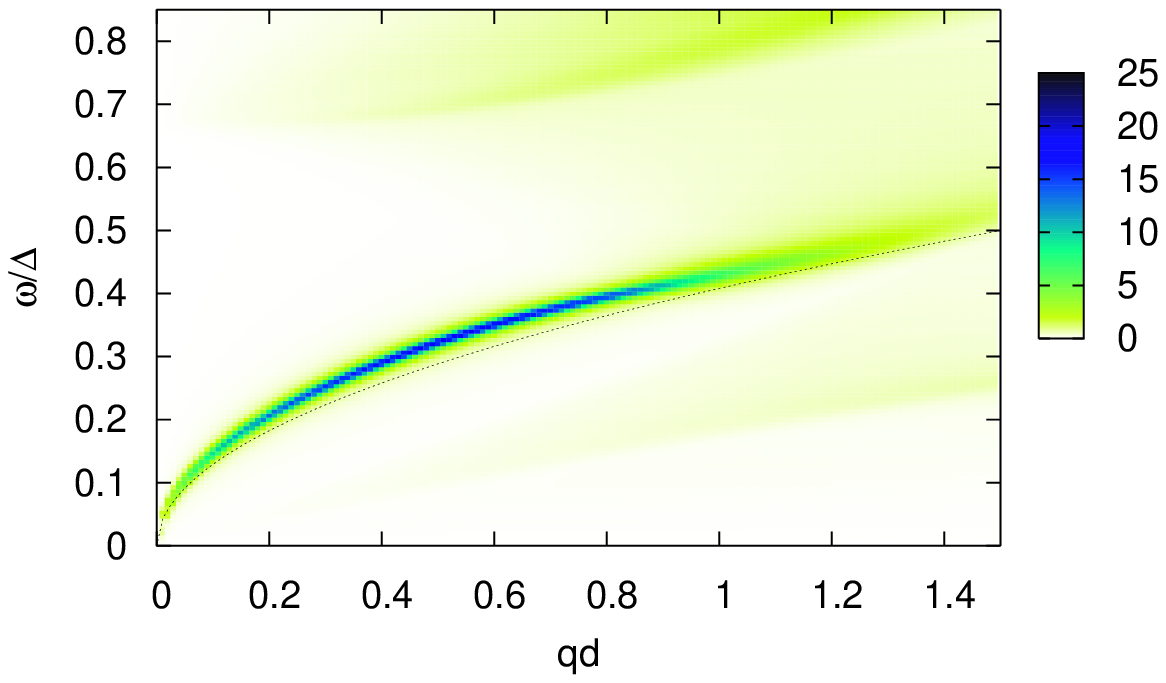} 
	\includegraphics[scale=0.83]{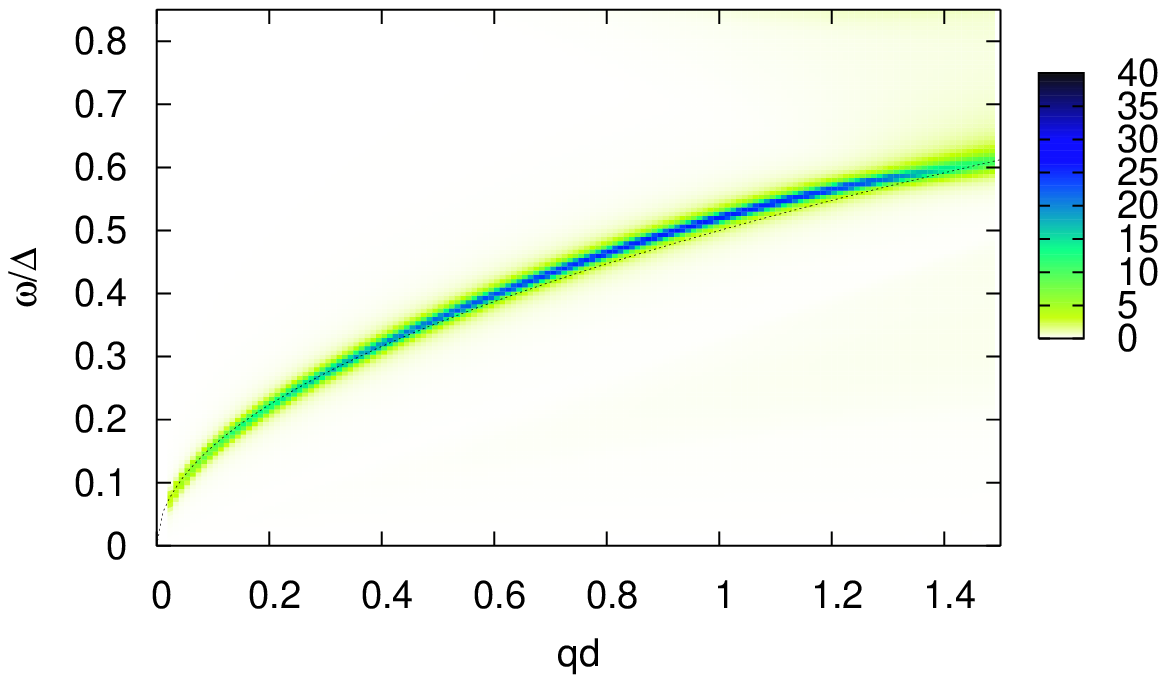}
	\includegraphics[scale=0.83]{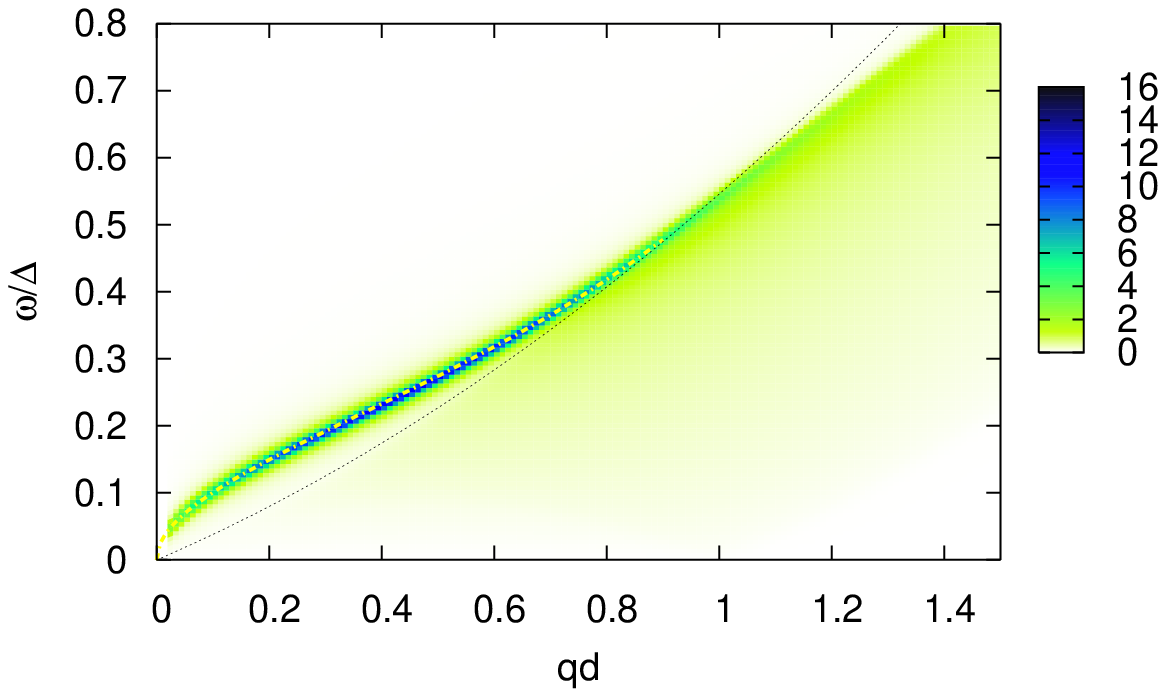}
\end{center}
\caption{(Color online) Plasmon dispersion for $q_x=0$ and different values of the Fermi energy $\mu$.
a) $\mu=0.33$, b) $\mu=0.5$, c) $\mu=-0.92$. In Fig. c) we compare with the plasmon dispersion of the
conventional (parabolic) 2D electron gas (dashed curve). The curve in c) is the plasmon dispersion of a 
conventional 2D electron gas of Eq. (\ref{conveg}), below the dotted line is the intraband single particle excitation area of a 2D electron gas.}
\label{energies}
\end{figure}
\begin{figure}[ht]
\begin{center}
 	\includegraphics[scale=0.83]{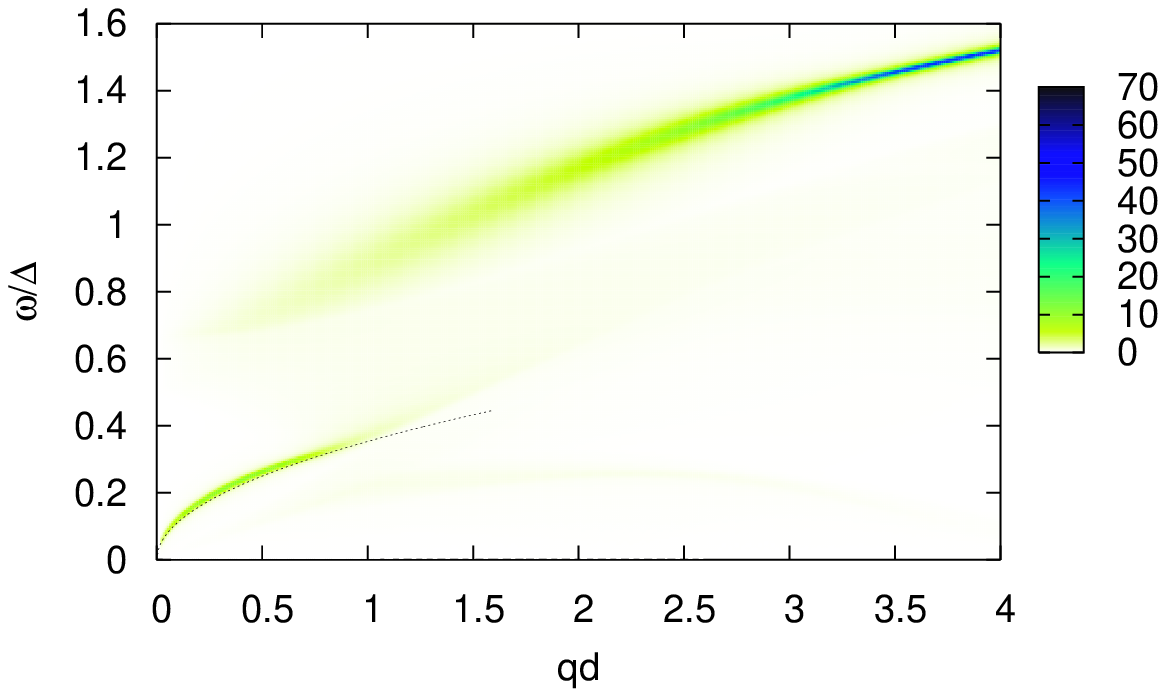}
\end{center}
\caption{(Color online) Plasmon dispersion for $\mu=0.25$ and $q_x=0$.}
\label{largeq}
\end{figure}

\section{Conclusion}
In this paper we have studied the dielectric function of graphene and the plasmon dispersion
in tight-binding approximation, and compared the results with similar calculations for a single Dirac cone.
There are differences due to the full energy spectrum of the electrons. It turns out that the damping of 
the plasmons is stronger, which indicates more scattering between plasmons and single electrons on the
honeycomb lattice. Moreover, the plasmon dispersion is not isotropic. Although the dispersion fits well the
sqare-root behavior $\omega_P\sim q^{1/2}$ at small $q$ values, there are substantial deviations for
larger values. Finally, a new branch appears in the plasmon dispersion for higher energies $\omega$ 
and larger wave vectors $q$ which needs to be subject of further investigations. Our results reflect a crossover from Dirac-like behavior to conventional electron gas behavior by changing the electron density with the help of a gate.

\section{Acknowledgements}
We are grateful to Timur Tudorovskiy for useful discussions. This work was supported by the Deutsche Forschungsgemeinschaft.

\end{document}